\documentstyle[aps, 12pt,psfig]{revtex}
                                                       
\begin{document}                                                                
                                                                                
\def \vsp {\vspace{.4cm}}
\def \vss {\vspace{.4cm}}                                                       
\def \vsl {\vspace{1cm}}                                                        
\def \fl {\flushleft}                                                           
\def \fr {\flushright}                                                          
\def \LAB {$A^{1/3} + B^{1/3}$ }
\def \LAC {$(N-1)/2 + (M-1)/2 $ }
\def \jps {$J/\Psi$'s }
\def \soc {$\sigma_{oc}$ }
\def \jp  {$J/\Psi$ }
\def \xf  {$x_F$ }
\def \pt  {$p_t$ }
\def \dsdn {$d\sqrt s/ dn$ }
\def \S {$\sqrt s$ }

\begin{center}
{\bf \jp Suppression and the Quark-Gluon Plasma}

\vss
S. Frankel and W. Frati

\vsp
Physics Department, University of Pennsylvania

\vsp   October, 1997

\vsl  \vsl \vsl

{ \bf Abstract }

\end{center}

      All measured Feynman \xf distributions of the ratio, R, of \jp
production
in nuclei relative to production on protons fall off with \xf.
They  show \cite{worldsci}
that
absorption of charmonium cannot be the only source of \jp suppression and
that energy loss of the constituents of the
incident proton prior to the \jp production,
                           because of the exponential \S
dependence of the charmonium cross section,
      should not     be neglected.
Including the effects of initial state  energy loss
we  find that the latest measured Pb-Pb \jp cross sections
     do not provide any evidence for  deconfinement.

\newpage

      Over a decade has passed since it was proposed that \jp suppression
in p-nucleus collisions was evidence for creation of a quark-gluon plasma.
\cite{Satz}  It was soon shown \cite{worldsci} from studies of the
Feynman x distributions of relative p-A and p-p production of the
\jp
that initial state and final
state interactions could account for the data. It is now claimed that
recent measurements show an anomalous depression in the total cross
section for J/Psi production in Pb-Pb interactions. No Feynman x
distributions of these data have been reported.

This note gives results of
 the full analysis taking into account both initial state
energy loss and final state \jp absorption.
Fig. 1 shows the data for
two of the many examples of the ratio, R, of the nuclear to
hadronic
\xf distributions of the cross
sections in pion and proton interactions.
\begin{figure}[t]
 \centerline{\psfig{figure=jpffig1,height=12cm}}
 \caption{}
\end{figure}
                     The pion data are from ref.
\cite{Badier} and proton data from ref. \cite{Alde}.

The fact that the ratio     is not independent of \xf rules out \xf
independent
absorption
as the {\it sole} mechanism for \jp suppression, and the difference
 in the suppression in pion- and
proton induced reactions mirrors     the experimentally measured
difference in energy loss in the initial state soft interactions.

(Theoretical fits, superimposed  on Fig 1., and reproduced from early
 work
\cite{zeitphys} \cite{UPS}, include
an energy loss per collision, \dsdn = 0.4. These fits showed
              the effect of a final
state \jp-nucleon  interaction with
a \jp + n $\rightarrow$ \jp +X  inelastic cross section of about
ten millibarns, and a final state energy loss of 10 \% per collision.
The fits  are only illustrative since trade-off between them can also 
account for the data. Fortunately,
the suppression in the total cross section is simpler to understand
             than that
seen in the \xf distributions since it is unaffected by the
inelastic scattering cross-section and energy loss
of charmonium in the final state, only by the initial state
energy loss and the 
open-charm absorption cross section.)

The energy loss of the particular nucleon
constituents
that finally form charmonium
cannot be calculated
because of the non-perturbative nature of the prior soft collisions.
One can only estimate the prior energy loss per collision,
\dsdn. The energy loss distributions for soft minimum bias collisions
can be determined from the experimental
event structures at a particular \S. The ISAJET program of Frank Paige
reproduces that data and allows for interpolations between different \S.
It allows one to
{\it estimate} the relative values of a and b in the
expression, \dsdn = a + b\S.,
This relationship has been well tested in studies of
low \pt interactions in nuclei \cite{nucphys} \cite{predicting}.
  Nevertheless for different
high \pt final states a and b should be taken as parameters that can be
extracted by studying data for widely varying A.

(Another equivalent method is to assign an
    energy loss to the quarks or gluons that are responsible for the
production of the particular final state that is being studied, e.g,
\jps or Drell-Yan pairs, again assuming some energy loss parameter.
This was the approach of the work of Gavin and Milana.\cite{Gavin}.
                                                               Their
ratio of energy loss for Drell-Yan relative to \jp production,
(4/9), is actually close to the values obtained in
refs.\cite{zeitphys},    \cite{UPS}, and \cite{suppressions}).

      It is worth reviewing some concepts of time evolution in the
initial state. For example,
in a p-p interaction resulting in the production of
a pion, the produced unclothed q-qbar pair
eventually picks up the gluons needed to form the final on-shell pion.
Because the pion has an extended structure the evolution of the bare pair
is traditionally believed to take a time approximately equal to r/c
where r is the pion dimension. However, in the case of {\it point} 
particles,
there is no finite size, and, as in electron-electron scattering, the
initial state quarks in a p-p collision
simply scatter, resulting only in a change in their momenta.
(Even if the quarks and gluons within a proton are not exact points, they
surely have a much smaller radius than a hadron
and will have a correspondingly shorter evolution time.)
Eventually the quarks will form a real on-shell pion but, before that
occurs, gluon constituents in the colliding protons
will have produced charmonium. Charmonium, the unclothed \jp,
will also take time to evolve into a real \jp  but it is the charmonium
state that actually interacts with other nucleons, causing its breakup
into open charm. (Of course even the $\Delta t \Delta E $ argument that
describes  the time evolution in free space will be affected if the
q-qbar pair is made, not in vacuum, but inside of nuclear matter.)

It is sometime argued that Drell-Yan production in nuclei shows that
there are no initial state energy losses that affect the Drell-Yan
cross-section and that, therefore, initial state interactions can be
entirely neglected. It is claimed that this is shown by the measured
$A^\alpha $
dependence of Drell-Yan production in nuclei. The data do show a value of
$\alpha$ slightly less than unity so this conclusion cannot be correct.
The suppression of the \jp relative to the suppression of Drell-Yan pairs
depends both on the energy loss and the {\it shape} of the \xf
 distribution. The \xf distributions for p-p interactions are quite
different in these two cases, varying  roughly as $(1-x_f)^n$ with n = 5.2
for the \jp and 2 for the Drell-Yan. This makes the energy loss effects
larger for the \jp. Also, in Drell-Yan,
  if the dimuon is made in the first
collision there is no energy loss,  so events can  appear at \xf = 1.
In this case there is no final state absorption so
the fraction of such events is just the Glauber probability of getting a
single collision which is $1/<n>$ where $<n>$ is the mean number of 
collisions. But, if there is energy loss, those dimuons will not populate
\xf = 1. Thus the suppression will turn over and
drop  to $1/<n>$  This will give larger
depressions at \xf =1 for heavy nuclei.
                                     One has to look, therefore, at the
high \xf region to learn about energy loss effects in Drell-Yan production
in nuclei. The variation of the suppression with \xf has been calculated
for a few cases in references  \cite{UPS} and \cite{dependence}.

 It should be
understood that the energy loss of the gluons involved in charmonium
production need not be the same as the energy loss of the quarks producing
Drell-Yan pairs. It will be the task of non-perturbative physics in the
future to calculate the relative energy losses. In our work we attempt to
use approximate values taken from low \pt experiments summarized in the
ISAJET model. \dsdn for Drell-Yan production can be determined from
the Drell-Yan data at large \xf while \dsdn for the \jp data is 
complicated
                          by the effects of final state absorption and
scattering.

To constrain the new analysis we have made use of the independent
determination of the
\jp absorption cross section from photoproduction since that reaction has
no initial state interactions. That value is  \soc = 6.6 +/- 2.2 mb,
as obtained in reference \cite{zeitphys}.

To calculate the \jp suppression we have again used a full Monte Carlo
calculation, generating nucleons according to the 3 parameter
Woods-Saxon distribution, $\rho(r) \ \frac{1}{ 1 + e^{(r-R_0)/\sigma} }$
 with  $R_0 = 1.19 A^{1/3} - 1.61 A^{-1/3}$, and $\sigma$ = .545.
 Protons are then scattered off the nucleons in
the nuclei, to count the number of nucleon collisions prior to the \jp
production, (which determines
the energy loss), and to count the number of
succeeding collisions of the charmonium,
(which determines the absorption).
This method removes the struck nucleon from the counting \cite{walet}.
(This differs from
calculations \cite{kharzeev}
         which integrate the paths through a continuous
nucleus matter distribution.
 Not taking into account the finite nucleon
size underestimates determination of an absorption cross section and may
account for differences in an extracted \soc.)

Unlike charmonium, produced in a high $Q^2$ reaction, the low $Q^2$
$q-\bar q$ pairs (e.g., off-shell pions), produced in the soft
 nucleon-nucleon collisions
have a longer evolution time. This is in fact the basis for the
difference between counting participants (``wounded nucleons'')
 in predicting minimum bias event
structures \cite{brody},
 rather than the number of scatters as needed for  high $Q^2$
charmonium production.
Thus pions produced in the final state
    should not contribute significantly to charmonium
absorption, are not included in the calculation, and, as we shall see,
are not needed to account for the new nucleus-nucleus data.
We also do not distinguish between charmonium absorption by  unstruck
nucleons, as in the p-A
case, and on absorption on previously struck protons,
as in the B-A case.

The parameterization\cite{Craigie}
 of the \jp production cross section,
proportional to $e^{-\gamma M/\sqrt s}$, ($\gamma M = 45$),
is used. It is responsible for the
\S dependence of the suppression, with $\sqrt{ s(n)} = \sqrt {s(0)} -
(d\sqrt s/dn) n$, where n is the number of prior collisions.

 \begin{figure}[t]
\centerline{\psfig{figure=jpffig2a.1,height=10cm}}
 \caption{}
  \end{figure}

      We show in Fig. 2
our calculations for both the mean \soc obtained from our prior
photoproduction analysis and a
somewhat  larger value, using our estimate of the energy loss, 
$d\sqrt s/dn = 0.5 + 0.018 \sqrt s$.
  One can
make small trade-offs of
  \soc with \dsdn to get other suitable fits to the 
data.  The filled points show the data for \S = 38.8 GeV
(circles), \S = 19.4 GeV
(triangles) and \S = 17.4 GeV
(squares) which are the results of different
experiments \cite{lourenco}.
 The open squares are the calculations for
\soc = 6.3 mb.,  while the open circles are for 7.9 mb.

In Fig. 2 we have drawn a straight line, obtained by fitting only the
   low higher  accuracy \LAB points, which
appear to show an exponential behavior. It extrapolates to a value well
above the value for  Pb-Pb. However there is no 
{\it theoretical} reason for an exponential
extrapolation procedure to be valid out to large  \LAB,
  even if absorption were the only
mechanism. In fact it is  shown \cite{suppressions}
 that, even with pure absorption,
               analytic calculations and the functional dependence of
prior and subsequent scatters as a function of \LAB should cause a small
{\it enhancement}
of R above experiment at very high \LAB.

The \jp suppression is a valuable tool for studying many features of
reactions that have simultaneously
both low $Q^2$ and high $Q^2$ interactions. We need
to know more about the time evolution of charmonium into the on-shell
\jp, the changes in nucleon structure functions resulting from the prior
soft interactions, and the possible variation of the \jp absorption 
cross section with outgoing
energy. This requires more data and analysis of photoproduction
as well as Drell-Yan production, especially the small, but not zero,
energy losses that occur in the latter reaction. The Drell-Yan \xf
distribution measurements
  should be extended  to \xf = 1,  to verify that $R(x_f = 1)$
depends only on the Glauber probability that the dimuon is made
in the {\it first} collision \cite{dependence}.

\vsp
Conclusion: Incorporating the presence of energy loss in the soft
interactions prior to \jp production, we find no 
evidence for effects of deconfinement.

\vsl

\centerline{ \bf Figure Captions }

Fig. 1  Two examples of the ratios of hadron-nucleus to hadron-proton
\jp cross sections vs. Feynman x. The fits are taken 
from ref. \cite{zeitphys} and described in this text.

\vss
Fig. 2  Comparison of \jp suppression ratios, R, for experiments at
different \S, plotted vs \LAB. Fits are shown
for two different values of \soc , showing agreement of
our theoretical predictions with the data for Pb-Pb collisions which
fall below a straight line {\it extrapolated} from the logarithmic plot
of lower \LAB  data.

\fr jpletted6sent  printed   \today
\end{document}